\documentclass[oldversion]{aa}

\usepackage{graphicx}
\usepackage{natbib}
\usepackage[varg]{txfonts}

\bibpunct{(}{)}{;}{a}{}{,}

\usepackage{txfonts}

\begin{document}

   \title{X-ray narrow line region variability as a geometry probe}
   \subtitle{The case of NGC 5548}

   \author{R.G. Detmers\inst{1}
          \and
          J.S. Kaastra\inst{1}\inst{,2}
          \and
          I.M. McHardy\inst{3}}

   \offprints{R.G.Detmers}

   \institute{SRON Netherlands Institute for Space Research, Sorbonnelaan 2, 3584 CA Utrecht, The Netherlands \email{r.g.detmers@sron.nl}
    \and
    Astronomical Institute, University of Utrecht, Postbus 80000, 3508 TA Utrecht, The Netherlands
    \and
    School of Physics and Astronomy, University of Southampton, Southampton SO17 1BJ, UK}     

   \date{}

 
  \abstract{We study the long time scale variability of the gas responsible for the X-ray narrow emission lines in the Seyfert 1 galaxy NGC 5548, in order to constrain the location and geometry of the emitting gas. Using X-ray spectra taken with the \textit{Chandra}$-$LETGS and HETGS instruments and with XMM$-$\textit{Newton} RGS and combining them with long-term monitoring observations of the \textit{Rossi X-ray Timing Explorer} (RXTE), we perform a correlation analysis in order to try constrain the time scale on which the narrow line emitting gas responds to variations of the continuum flux. With the inclusion of the 2007 \textit{Chandra}$-$LETGS observation we have an additional observation at an historically low flux level. We conclude that the NLR in NGC 5548 is in the form of an ionization cone, compact in size, and located between 1 and 15 pc from the central source, depending on the exact geometry of the NLR.}

  \keywords{active galactic nuclei --
                variability --
                emission lines -- individual: NGC 5548 -- galaxies: Seyfert}
  \maketitle
%

\section{Introduction}			\label{intro}

Variability in Active Galactic Nuclei (AGN) is currently one of the best methods to study the properties of the gas surrounding super-massive black holes (SMBH). Whether it is the optical Broad-Line Region (BLR) studied by reverberation methods \citep[see e.g.][]{Peterson04} or the warm absorber as studied in the UV \citep[for example][]{Kraemer06} and X-ray bands \citep[see e.g.][]{Netzer03,Krongold07}, variability allows us to investigate the physical properties and location of the gas without directly resolving the inner regions of AGN.

The narrow line region (NLR) is the region responsible for the narrow forbidden and coronal lines as observed by optical and UV observations \citep[see e.g.][]{Kraemer98}. The density of this gas is much lower ($10^{10-12}$ m$^{-3}$) than that of the BLR ($10^{16}$ m$^{-3}$) and the width of the lines is also much smaller, less than 1000 km $\mathrm{s^{-1}}$. The exact origin of this gas is unknown.  
The X-ray narrow line region is well known from the observations of various Seyfert 2 galaxies \citep{Kinkhabwala02, Bianchi06} which show that it consists of photoionized gas in a cone-like geometry of sizes on the order of 100 pc. This gas may be the 'warm absorber' gas seen in emission, although the emission lines show a blueshift in some sources, e.g. NGC 1068 \citep{Kinkhabwala02}, but not in others, e.g. Mrk 3 \citep{Bianchi05}. 

In Seyfert 1 galaxies it is much harder, with respect to the emission line spectrum observed in Seyfert 2 galaxies, to detect the different narrow emission lines due to the bright continuum which washes out the weaker lines. The strongest line visible in a typical Seyfert 1 high-resolution X-ray spectrum is the \ion{O}{vii} forbidden emission line, although other lines are also detected depending on the quality of the spectrum (among others \ion{Ne}{ix} f and \ion{C}{v} f). No significant blueshift is detected in the emission lines of NGC 5548 \citep{Kaastra02b}, while NGC 3783 shows evidence of redshifted emission lines, which could be an indication of a P-Cygni like wind \citep{Kaspi02,Behar03}. 

Recent work on Mrk 335 \citep{Longinotti08}, done while the source was in a low state, revealed the presence of emission lines, which the authors place at the location of the optical BLR clouds, which indicates that the emitting region is very compact in size. This a good example of why studying the X-ray NLR in Seyfert 1 galaxies is important. Our understanding of the location and geometry of the X-ray NLR is lacking at the moment, so constraining these is important for determining the origin of the NLR and whether it is connected to the warm absorber. The best method to do this in Type 1 AGN is to take advantage of variability and combine that with high-resolution X-ray spectroscopy. 
 
NGC 5548 is one of the best studied AGN, with optical observations spanning 30 years \citep{Sergeev07}, seven high-resolution X-ray observations and 11 years of monitoring with the \textit{RXTE} satellite. This makes it one of the best AGN for studying the intrinsic variability of the X-ray NLR. 
\citet{Detmers08} (hereafter D08) previously constrained the location of the narrow emission line region to within 1 pc of the central source (assuming a spherical NLR). This was based on the line flux change between 2002 and 2005, yielding an upper limit to the size of the emitting region of 3 lightyears. With the addition of the new 2007 observation we decided to perform a correlation analysis between the emission line fluxes and the continuum flux in order to see if the upper limit of 1 pc could be further refined and how it depends on the geometry of the NLR.
We therefore present results derived from seven high-resolution X-ray observations taken by the XMM$-$\textit{Newton} RGS and the \textit{Chandra}$-$LETGS and HETGS instruments. These observations give us precise flux measurements for the narrow emission lines. The continuum flux history of NGC 5548 is determined using the \textit{RXTE} observations, which span the period of 1996 to 2007. NGC 5548 has dropped in flux, on average, by a factor of two in the 2 $-$ 10 keV band since 2002. This low "state" lasts up until the last measurement in September 2007. The LETGS observation of 2007 caught the source at the lowest flux level observed so far, providing us with two observations when NGC 5548 was at a low flux level (2005 being the other observation). The long \textit{RXTE} lightcurve gives us the opportunity to search for correlations between the strength of the narrow emission lines and the continuum flux. 
We discuss the observations used and the data reduction briefly in Sect. \ref{data}. Sect. \ref{analysis} contains the correlation analysis for the continuum and the emission lines. In Sect. \ref{geo} we derive the geometry of the NLR. We discuss our results in Sect. \ref{disc} and present our conclusions in Sect. \ref{conclusions}.  

 \section{Data reduction}				\label{data}
   
We have used data from all 7 observations of NGC 5548 as taken with \textit{Chandra}$-$LETGS and HETGS and with XMM$-$\textit{Newton} RGS. Table \ref{tab:data} shows the date and the instruments used for each observation. Except for the 2007 LETGS data, all data have been previously published, as indicated in Table \ref{tab:data}. The LETGS data were reduced as described in D08 and we have followed the same procedures for fitting the data and error analysis. Briefly, the data were reduced using the standard CXC pipeline up until the level 1.5 event files. After that we follow an independent procedure up to the level 2 event files. 
For the two RGS observations we have obtained the data using the public archive and used SAS version 8.01 to reduce the data. We use C-statistics for fitting the spectrum. Spectral fitting is done using the SPEX package \citep{Kaastra96}.

\begin{table}
\caption{High resolution X-ray observations of NGC 5548.  }             
\label{tab:data}      
\begin{center}                          
\begin{tabular}{l@{\,}c@{\,}c}       
\hline\hline                 

Year (month) & Instrument & Reference\footnote{}  \\    

\hline                        

 1999 (Dec) &LETGS & 1    \\
 2000 (Feb) &HETGS & 1 + 2  \\
 2000 (Dec) &RGS & 6 \\
 2001 (Jul) &RGS & 3 + 6 \\
 2002 (Jan) &LETGS + HETGS & 4 \\   
 2005 (Apr) &LETGS & 5 \\
 2007 (Aug) &LETGS & 6 \\ 

\hline
\end{tabular}
\end{center}
\footnotesize{$^1$ References for observations: (1) \citet{Kaastra02b}; (2) \citet{Yaqoob01}; 
(3) \citet{Steenbrugge03}; (4) \citet{Steenbrugge05}; (5) D08; (6) Present work.} 
\end{table}
 
\section{Data analysis}   	\label{analysis}

For all observations we used the following method to obtain the \ion{O}{vii} f line parameters. 
We model the spectrum with a power-law and blackbody component, absorbed by three photoionized components that are modelled using the \textit{xabs} model of SPEX. Also the cosmological redshift and Galactic absorption are taken into account. We model the \ion{O}{vii} f line with a Gaussian line, which we put outside the range of the warm absorber gas, so it is only affected by the cosmological redshift and the neutral interstellar absorption. This is the same model which was used in D08 to model the spectrum.

The 2007 LETGS observation was taken when the source was at a historically low flux level (F$\mathrm{_{2-10\,keV}}$ = 8.5 $\times$ $10^{-15}$ W $\mathrm{m^{-2}}$), which unfortunately prevented any detailed analysis of the warm absorber due to low statistics. The detected \ion{O}{vii} f line is listed in Table \ref{tab:lines} along with previous detections in the other observations. The \ion{O}{vii} f line flux is different with respect to D08 for the two RGS measures. This is due to the fact that the analysis in D08 was a preliminary one, which was done using the online RGS BIRD\footnote{http://xmm.esac.esa.int/BiRD} archive. The new fluxes were obtained by using the above method.

\begin{table}
\caption{The unabsorbed flux in ph $\mathrm{m^{-2}}$ $\mathrm{s^{-1}}$ for the \ion{O}{vii} forbidden emission line for all high resolution spectral observations of NGC 5548. All errors are calculated at a 68 \% confidence level. }             
\label{tab:lines}      
\begin{center}                          
\begin{tabular}{l c}        
\hline\hline                 

Year (month) & \ion{O}{vii}  \\ 

\hline                        
1999 (Dec) & 0.81$\pm$0.16 \\
2000 (Feb) & 0.82$\pm$0.18 \\
2000 (Dec) & 1.3$\pm$0.2 \\
2001 (Jul)  & 1.1$\pm$0.1  \\
2002 (Jan) & 0.75$\pm$0.07 \\     
2005 (Apr) & 0.35$\pm$0.06  \\  
2007 (Aug) & 0.27$\pm$0.06 \\ 

\hline
\end{tabular}
\end{center}
\end{table} 

\subsection{Method and Results} \label{methods}

Reverberation mapping methods use the cross-correlation of the continuum light curve with the emission line lightcurve. However due to the sparse sampling of the emission line data we have chosen another method to address this specific problem.

The observed emission line flux does not respond instantaneously to the continuum flux changes, but due to delay effects in the source region the signal will be smeared out over a time $\tau_\mathrm{{var}}$. Therefore we calculated the average continuum flux before each spectral observation at each of a series of time scales. For example, at the 16 days time scale the average flux between the time of an observation and 16 days before that observation is used. We assume that any variability at $\tau$ $\ll$ $\tau_\mathrm{{var}}$ is washed out. So we have seven $\ion{O}{vii}$ f snapshots and a corresponding average continuum flux for different timescales.

Then we checked for possible correlations at different time scales, the shortest time scale being 7 days and the longest 1280 days. These timescales were chosen because below 7 days the \textit{RXTE} lightcurve does not have enough data points in most of the bins. The upper timescale was chosen to correspond to approximately 3 years, which is the upper limit to the variability timescale (the time between the 2002 and 2005 observations). For the longest timescales, the continuum fluxes derived for some observations (basically the first five) are no longer independent of each other, due to overlap. For example, for the 1999(Dec) and 2000(Feb) observations the average continuum flux on a timescale larger than a few months will be the same, because the \textit{RXTE} data points are the same. 

\begin{figure}[tbp]
    \includegraphics[angle= -90,width=9cm]{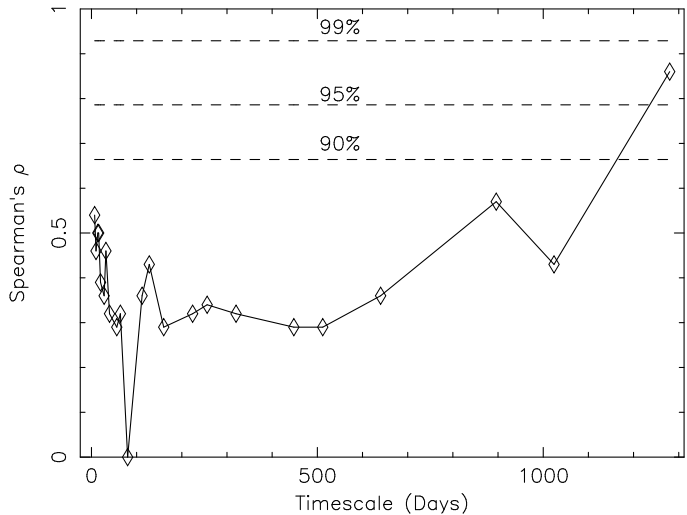}
    \caption{\label{fig:spearman} Spearman's $\rho$ for different timescales. The timescales tested are indicated by the diamonds, the three dashed lines are the 90, 95 and 99 $\%$ probability curves for the correlation degree. }
\end{figure}

In order to see if there is any correlation between the \ion{O}{vii} f fluxes and the average continuum flux on a certain timescale, we perform a Spearman rank coefficient analysis for each timescale. The results are shown in Fig. \ref{fig:spearman}. For all timescales the correlations are not significant. Only at the 1280 day timescale do we find a somewhat significant (95 $\%$) correlation, but this method does not include the errors on the average fluxes.

\subsection{Short term variability} \label{short}

\begin{figure}[tbp]
   \includegraphics[angle= -90,width=9cm]{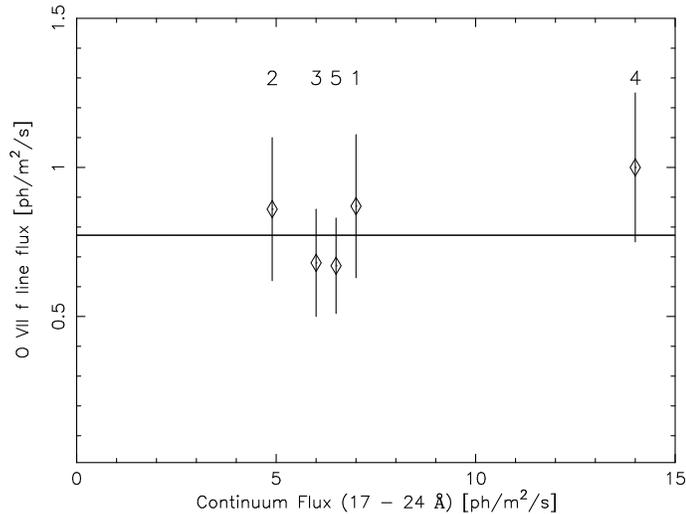}
   \caption{\label{fig:short}
   The short time scale variability of the \ion{O}{vii} f emission line. Labels indicate the time sequence of the data points. The solid line shows the average line flux of all five observations.}
\end{figure}

For one of the spectral observations we have a whole week of observing time, namely the 2002 LETGS + HETGS observation \citep{Kaastra04,Steenbrugge05}. As discussed in \citet{Kaastra04}, the source experienced a large flare during the LETGS observation, which allows us to check the short term variability of the \ion{O}{vii} f line during that week. During different time segments before, during and after the flare \citep{Kaastra04}, we fitted a powerlaw between 17 and 24 $\AA$ and used a Gaussian line to fit the \ion{O}{vii} f emission line. Fig. \ref{fig:short} shows the 5 different measurements as well as the fitted average line flux (solid line). All measurements are consistent with the line flux being constant. Despite the factor of 3 variation in continuum flux, the line did not change significantly (less than 25 \%).
This gives us a lower limit of 3 days for the variability time scale of the emission line region, which is consistent with the lower limit derived from the width of the line (D08).

\section{Location and geometry of the NLR} 	\label{geo}

There are four observational quantities based on the \ion{O}{vii} f line properties that can be used to constrain the location of the X-ray NLR:
\begin{itemize}
\item The variability time $\tau_{\mathrm{var}}$
\item The velocity width $V_{\mathrm{FWHM}}$
\item The emission measure $Y$
\item The ratio between the narrow $\ion{O}{vii}$ f and $\ion{O}{viii}$ Ly$\alpha$ line flux.
\end{itemize}

\begin{figure}[tbp]
   \includegraphics[width=9cm]{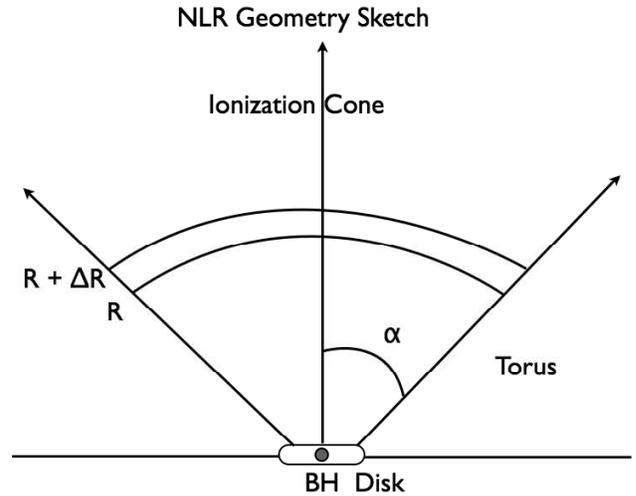}
   \caption{\label{fig:sketch}
   A sketch of the geometry of the X-ray NLR region in NGC 5548. The ionization cone has a half-opening angle $\alpha$ and is confined by the torus. The X-ray NLR gas is located in a slab with thickness $\Delta$\,R at a distance R from the central black hole.   }
\end{figure}

In order to constrain the location of the NLR, we make the following assumptions. First, we assume that the emission arises from a cone-shaped region, between radii $R$ and $R$\,+\,$\Delta\,R$. The cone has a half opening angle $\alpha$. 
Secondly we assume that the ionization parameter $\xi$ can be determined from the ratio of the narrow $\ion{O}{vii}$ f and $\ion{O}{viii}$ Ly$\alpha$ lines. The fluxes for these lines are taken from the February 2000 HETGS observation. We therefore make the assumption that the ionization state of the gas is similar in January 2002, since the $\ion{O}{vii}$ f line fluxes are similar in 2000 and 2002. From the observed line ratio and using XSTAR photoionization models, we determine that both lines ($\ion{O}{viii}$ and $\ion{O}{vii}$) can be produced by gas with log $\xi$ = 1.2 $\pm$ $0.2$.
Further we assume that the density drops rapidly with distance from the center (for instance $n$ $\sim$ $R^{-2}$). There are two reasons for making this assumption. First, for a spherical outflow the mass outflow is $\dot{M}$ $\sim$ $n\,R^{2}\,v$. Most outflow models do not predict a broad range of values for $v$, implying approximately $n$ $\sim$ $R^{-2}$.
The other argument is based on studies of Seyfert 2 ionization cones by \citet{Bianchi06}. They show that density laws similar to $n$ $\sim$ $R^{-2}$ are preferred in order to explain the soft X-ray emission and the [\ion{O}{iii}] profiles.
Since the emission scales as $n^{2}$\,$V$, this implies that most of the emission comes from a limited range of radii, even when $R$ $\ll$ $R$\,+\,$\Delta\,R$. Thus in general we can approximate the thickness of the region $\Delta\,R$ $\ll$ $R$, or in the worst case $\Delta\,R$ and $R$ are of the same order of magnitude. Finally we take the inclination angle $i$ of the cone to be zero, i.e. we look straight into the cone. This approximation is sufficiently accurate for our order of magnitude estimates.
Fig. \ref{fig:sketch} shows a sketch of the geometry of the X-ray NLR in NGC 5548.

\subsection{Analysis}	\label{ana}

There are several basic equations which are needed in order to constrain the location and geometry of the NLR. 
First the variability time scale $\tau_\mathrm{var}$ which is the sum of the recombination time scale and the light travel time:
\begin{equation}		\label{var}
\centering
\tau_\mathrm{var} = \tau_\mathrm{lt} + \tau_\mathrm{rec}
\end{equation}

Secondly, the ionization parameter $\xi$ is determined by the ratio of the ionizing flux and the density of the gas \citep{Tarter69}:
\begin{equation}		\label{xi}
\centering
      \xi = \frac{L}{n\,R^{2}} .  
\end{equation}
Here $L$ is the 1 $-$ 1000 Rydberg luminosity, for which we use the value of the 2002 observation (see Table \ref{tab:parameters}), $R$ is the distance to the gas and $n$ is the electron density.

Third, the recombination time $\tau_{\mathrm{rec}}$ scales inversely with the density $n$ of the gas and depends on the recombination rate of the specific ion \citep{Krolik95,Bottorff00}:
\begin{equation}		\label{trec}
\centering
      \tau_{\mathrm{rec}} = \left({\alpha_{\mathrm{r}}(X_{i})\,n\,(X_{i}) \left[\frac{f(X_{i+1})}{f(X_{i})} - \frac{\alpha_{\mathrm{r}}(X_{i-1})}{\alpha_{\mathrm{r}}(X_{i})}\right]}\right)^{-1} \equiv a\,(T,\xi)\,n^{-1}
\end{equation}
where $\alpha_{\mathrm{r}}(X_{i})$ is the recombination rate from ion $X_{i+1}$ to ion $X_{i}$ and $f(X_{i})$ is the fraction of element $X$ in ionization state $i$.   

The light travel time $\tau_{\mathrm{lt}}$ depends on the opening angle $\alpha$ in the following way, where we assume that the inclination angle $i$ of the cone with respect to our line of sight is zero, i.e. we are looking straight into the cone:
\begin{equation}		\label{tlt}
\centering
      \tau_{\mathrm{lt}} = \frac{\frac{1}{c}\int_{0}^{2\pi}\int_{0}^{\alpha}\int_{R}^{R\,+\,\Delta\,R}r(1-\cos\theta)r^{2}\sin\theta\,dr\,d\theta\,d\phi}{\int_{0}^{2\pi}\int_{0}^{\alpha}\int_{R}^{R\,+\,\Delta\,R}r^{2}\sin\theta\,dr\,d\theta\,d\phi}      .
\end{equation}
Which results in the following expression for the light travel time in the ionization cone:
\begin{equation}		\label{tlt}
\centering
      \tau_{\mathrm{lt}} = \frac{R}{c}\left(1 - 0.5\left(\frac{{\sin}^{2}\,\alpha}{1 - \cos\,\alpha}\right)\right)      .
\end{equation}

The emission measure $Y$ depends on the hydrogen density $n$ and the emitting volume $V$ as
\begin{equation}		\label{em}
\centering
Y  = 1.2 n^{2} V .
\end{equation}
Where the factor of 1.2 is a consequence of $n_{\mathrm{h}}$ = 0.85 $n_{\mathrm{e}}$. 
The volume $V$ depends on the half-opening angle $\alpha$ in the following way, with the assumption that $\Delta\,R$ $\ll$ $R$ (see discussion Sect. \ref{disc}):
\begin{equation}		\label{vol}
\centering
V \simeq 2\pi\,(1-\cos\alpha)f\frac{\Delta\,R}{R}R^{3} ,
\end{equation} 
with $f$ the filling factor of the emitting gas. 

If we assume that the NLR gas is moving in random Keplerian orbits and has an isotropic velocity distribution, we can determine the minimum distance from the upper limit on the linewidth of the \ion{O}{vii} f emission line using the following equation \citep{Netzer90}:
\begin{equation}		\label{fwhm}
\centering
V_{\mathrm{FWHM}}  = \left(\frac{4 G M }{3 R}\right)^{0.5} ,
\end{equation}
with $M$ the mass of the supermassive black hole and $V_{\mathrm{FWHM}}$ the FWHM of the line.

We use the observed luminosity of the \ion{O}{vii} f line ($L_{\mathrm{\ion{O}{vii}\,f}}$ = 5 $\times$ 10$^{33}$ W) to determine $Y$. This is done using an XSTAR run with log $\xi$ = 1.2, $L$ = $6.6 \times 10^{36}$ W and a temperature $T$ = 35\,000 K \citep{Steenbrugge05}. We obtain $Y$ = 4.3 $\times$ $10^{70}$ m$^{-3}$. 

\begin{table}
\caption{Important parameters of NGC 5548.  }             
\label{tab:parameters}      
\begin{center}                          
\begin{tabular}{c c c c}       
\hline\hline                 

Parameter & Description & Value &Reference\footnote{}  \\    

\hline                        

$L$ & Luminosity (1$-$1000 Ryd) & 6.6 $\times$ $10^{36}$ W & 1 \\
log $\xi$ & Ionization parameter & 1.2 & 1 \\
$a$  & Parameter defined in Eq. \ref{trec} &  1.64 $\times$ 10$^{17}$ m$^{-3}$ s  & 1 \\   
$M_{\mathrm{BH}}$  & Black hole mass &  6.54 $\times$ $10^{7}$ $\mathrm{M_{\odot}}$ & 2\\
$Y$ & Emission measure &   4.3 $\times$ $10^{70} $ m$^{-3}$ & 1 \\ 

\hline
\end{tabular}
\end{center}
\footnotesize{$^3$ References for parameter values: (1) Present Work; (2) \citet{Bentz07}} 
\end{table}

Inserting the observed parameter values as listed in Table \ref{tab:parameters} into (\ref{xi}) - (\ref{var}), these equations can be rewritten as functions of $R$ and $\alpha$ only:
\begin{itemize}
\item $n\,(\mathrm{m}^{-3})$ = 4.4 $\times\, 10^{11}$ $R^{-2}$
\item $\tau_{\mathrm{rec}}$\,(yr) = 0.012 $R^{2}$
\item $\tau_{\mathrm{lt}}$\,(yr) = 3.3 $\left(1 - 0.5\left(\frac{{\sin}^{2}\,\alpha}{1 - \cos\,\alpha}\right)\right)$ $R$
\item $V\,(\mathrm{pc}^{3})$ = 0.0064 $R^{4}$
\item $f$ $\frac{\Delta\,R}{R}$ = 1.02 $\times\, 10^{-3}\,\frac{R}{(1 - \cos\alpha)}$
\item $V_{\mathrm{FWHM}}\,(\mathrm{km}\, \mathrm{s^{-1}})$ = 610 $R^{-0.5}$
\item $\tau_{\mathrm{var}}\,(\mathrm{yr})$ = 3.3 $R$ $\left(1 - 0.5\left(\frac{{\sin}^{2}\,\alpha}{1 - \cos\,\alpha}\right)\right)$ + 0.012 $R^{2}$
\end{itemize}
Where $R$ is in pc in these expressions. 

The limits on the three observational quantities that can be used to constrain the location of the X-ray NLR are:
\begin{itemize}
\item  $V_{\mathrm{FWHM}}$ $\le$ 560 km $\mathrm{s^{-1}}$ (from Gaussian fit to emission line)
\item  $f$ $\frac{\Delta\,R}{R}$ $\le$ 1 (by definition due to density assumption)
\item  $\tau_{\mathrm{var}}$ $\le$ 3 years (see D08)
\end{itemize}

\begin{figure}[tbp]
    \includegraphics[angle= -90,width=9cm]{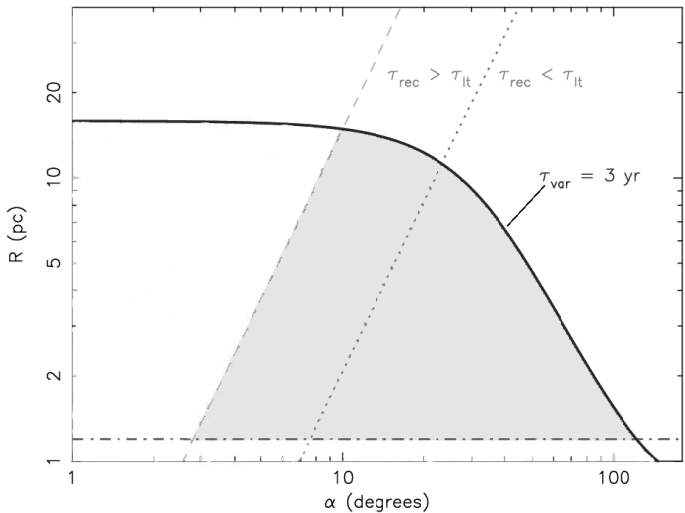}
    \caption{\label{fig:rlim1}
    The geometry of the NLR, based on an ionization cone model with half opening angle $\alpha$. The dotted-dashed line on the bottom is the lower limit to the distance as determined from the width of the \ion{O}{vii}\,f line. The left dashed line is the limit on the filling factor of the emitting gas (Eq. \ref{vol}). The thick solid line is the upper limit on the distance as derived from the variability time scale. The gray area indicates the allowed range for $R$ and $\alpha$.  The dotted line on the right indicates where $\tau_{\mathrm{lt}}$ = $\tau_{\mathrm{rec}}$. Due to the uncertainty in $\xi$, there is a typical formal uncertainty of 30 $\%$ in the $R$ value of the solid curve.}
\end{figure}

From $V_{\mathrm{FWHM}}$ we obtain a lower limit of $R$ = 1.2 pc. The $f\,\frac{\Delta\,R}{R}$ constraint leads to an upper limit of $R$ $\le\,2041\,(1 - \cos\alpha)$ pc. The limit on $R$ based on $\tau_{\mathrm{var}}$ is a quadratic equation which depends on the opening angle $\alpha$; it is shown in Fig. \ref{fig:rlim1}, together with all other constraints on $R$. The gray area shows the allowed geometry of the emitting gas, based on the above three constraints. These limits can in principal be tightened using additional information based on general properties of ionization cones in AGN. 

We can further constrain the location of the NLR if we assume that the warm absorber and the narrow line emitting gas are one and the same. The average column density can be written as
\begin{equation}		\label{col}
\centering
N_{H}  = n\,R\,f\,\frac{\Delta\,R}{R} .
\end{equation} 
Using the expression for $f$ $\frac{\Delta\,R}{R}$ obtained earlier ( (\ref{em}) and (\ref{vol}) ) we can rewrite (\ref{col}) into
\begin{equation}		\label{colangle}
\centering
N_{H} = \frac{Y\,\xi}{2.4\,\pi\,L\,(1\,-\,\cos\alpha)}
\end{equation}
For $\alpha$ between 3 degrees and 120 degrees (see Fig. \ref{fig:rlim1}), we find $N_{H}$ values for the emitting gas between 10$^{28}$ and 10$^{25}$ m$^{-2}$ respectively. By comparing this column density to the observed absorption column density we can determine if the NLR gas and the warm absorber gas could be the same. The observed column density of gas with log $\xi$ = 1.2 is 5 $\times$ 10$^{24}$ $\mathrm{m^{-2}}$ \citep[see fig. 5.4,][]{Steenbrugge05}. So given the uncertainties (factor $\sim$ 2$-$3) attached to the above method, we conclude that the warm absorber gas and the NLR gas could be one and the same, especially if we take into account that the covering factor $f$ is likely less than one (only 50$\%$ of the Seyfert 1 galaxies have a warm absorber.)

If we make the assumption that the gas that produces the $\ion{O}{vi}$ narrow lines also produces the X-ray narrow lines, we get an immediate estimate of the distance from the FWHM of the $\ion{O}{vi}$ lines, since they are resolved in the UV. Depending on the assumed model \citep[covered or uncovered NLR, see][]{Brotherton02}, the FWHM is either 432 $\pm$ 12 km $\mathrm{s^{-1}}$ or 658 $\pm$ 9 km $\mathrm{s^{-1}}$.
These values lead to estimates of $R$ of 2 pc or 0.9 pc, respectively.

From the minimum value for $R$ of 1.2 pc, it follows that the recombination time scale of the \ion{O}{vii} f emitting gas is at least 50 days. However from the lack of response of the \ion{O}{vii} f line to the continuum flux on different timescales we can in principle refine the lower limit of response timescale $\tau_{\mathrm{var}}$. Even though the continuum for the first 5 observations can vary a factor 1.5 between each observation up to a timescale of 600 days, the \ion{O}{vii} f line flux is consistent with being constant throughout these five observations. However the line flux for the two RGS observation is a factor of 1.5 higher than the others, if the errors are not considered. This is consistent with the picture that there can be small variations in the line flux on smaller timescales than 3 years, but that a large drop in average continuum flux for a long period is needed for the \ion{O}{vii} f line to respond significantly (like in the last two observations in 2005 and 2007). There is also the question of whether the gas is in ionization equilibrium. The continuum of NGC 5548 varies on time scales as fast as $\sim$ 1 day, with a relative amplitude of order unity (see also Sect. \ref{short} ). Therefore, strictly speaking, the gas cannot be in ionization equilibrium. However, it can be considered to be in quasi-equilibrium with the average continuum $\sim$ $\tau_{\mathrm{rec}}$ earlier.

\section{Discussion} \label{disc}

From Fig. \ref{fig:rlim1} we can already rule out several options for the geometry of the NLR. A spherical NLR is not possible, since the opening angle $\alpha$ is 180 degrees in that case. The upper limit on $\tau_{\mathrm{var}}$ along with the lower limit from $V_{\mathrm{FWHM}}$ rule out a half opening angle larger than $\sim$ 100 degrees, but anything larger than 90 degrees is unlikely, considering that the accretion disk is located at $\alpha$ = 90 degrees and thus blocks the line of sight to the other side. 
Therefore the most likely geometry for the NLR is an ionization cone, as observed in Seyfert 2 galaxies. The half opening angle of these ionization cones can typically vary between 20 and 45 degrees \citep{Schmitt01}.  Applying this additional constraint to the location of the NLR, we find that $R$ is between 1.2 and 12 pc.

The above constraints on the location have been obtained with the assumption that $n$ $\sim$ $R^{-2}$. This assumption is based on a constant mass outflow rate and on Seyfert 2 ionization cone studies \citep{Bianchi06}. Since we are not sure if the NLR is really outflowing, the second argument is the strongest indication that this assumption is correct as it is based on photoionization studies by comparing the optical [\ion{O}{III}] emitting region to the soft X-ray emitting region. They conclude that constant density models are excluded and that a steep density profile ($n$ $\sim$ $r^{-2}$) is preferred. Another study of six Seyfert 1 galaxies in the optical show different results however for the density slopes \citep{Bennert06}. Some Seyfert 1 galaxies show a slope $\sim$ $r^{-1}$, others a slope $\sim$ $r^{-2}$. When compared to Seyfert 2 galaxies in their sample, the Seyfert 1's on average show a steeper slope, but given the fact that there are only six sources, the scatter for individual sources can be quite large. Also it is not certain that the extended NLR that they probe (scales of $\sim$ 100's pc) is the same gas as the high-ionization X-ray gas we detect in NGC 5548. A better method to investigate the NLR properties would be using a grid of photoionization models for NGC 5548, but given the fact that we only have two lines in the X-ray and two lines in the UV to model, this would be too detailed for the current data we have. A future article where this analysis can be extended to include other Type 1 AGN would be better suited for this level of detail.

The constraint on the NLR distance in NGC 5548 places it at a much greater distance than \citet{Longinotti08} obtained in Mrk 335. However the line ratios of the oxygen triplet are different in Mrk 335 than in NGC 5548. The intercombination line is stronger than the forbidden line in Mrk 335, while in NGC 5548 this clearly is not the case. The line widths they find for the emission lines are also a factor of $\sim$ 3 greater than those of the \ion{O}{vii} f in NGC 5548. These two clues tell us that the emitting gas most likely is not the same in both sources. The gas in Mrk 335, has much more in common with the BLR (high density, large velocity width) than the NLR which we observe in NGC 5548.

From the observed warm absorber column density, we conclude that the warm absorber and the NLR gas could be the same given the uncertainties on the column density of the absorber ($\sim$ 30$\%$). However, the connection between absorber and emitter is still uncertain at this moment and will be hard to establish without first independently establish the geometry of both the absorbing and the emitting gas.

The connection between the narrow UV line emitting gas and the X-ray emitting gas is speculative at best at this moment, although the FWHM of both systems is similar and both the UV and X-ray line ratios can be produced by gas with a single ionization parameter (log $\xi$ = 0.9). A more detailed photoionization model, which takes into account the warm absorber and the line emission in both UV and X-ray, with the correct geometry and outflow velocities and time-dependent ionization effects, would be invaluable in comparing the UV and X-ray spectral information from the observations to the model predictions. However such a work is beyond the scope of this paper, since its main purpose is demonstrating the usage of variability as a geometry probe of the X-ray NLR.
Recently, \citet{Crenshaw09} have studied high-resolution UV spectra of NGC 5548, including one in an extremely low flux state in February 2004. This low flux spectrum reveals for the first time the presence of an intermediate line region (ILR) in NGC 5548. The \ion{C}{iv} emission line of this component has a FWHM of 700 km $\mathrm{s^{-1}}$ and is located at $\sim$ 1 pc from the nucleus. \citet{Crenshaw09} model this gas with $U$ = $10^{-1.5}$ (log\,$\xi$ = 0.1), $n$ = $10^{13}$ m$^{-3}$, a column density of 3 $\times\, 10^{25}$ m$^{-2}$ and a global covering factor of 0.06. The FWHM and location of the \ion{O}{vii} f line allow for the possibility that the X-ray emission comes from the same ILR region or only slightly further out. The main difference is that the X-ray emitting region has a $\sim$ 100 times lower density and a $\sim$ 10 times higher ionization parameter. This would be consistent with the UV emission coming from dense clumps that are embedded in a low density hotter gas which produces the X-ray emission. 
The \ion{O}{vii} f line centroid as observed by the LETGS agrees with the surrounding continuum to within 0.217$^{\prime\prime}$, which corresponds to 106 pc at the distance of NGC 5548 \citep{Kaastra03}, giving a independent upper limit to the size of the NLR. 

We therefore conclude that the NLR in NGC 5548 is compact in size, is in the form of clouds with a small covering factor or a narrow stream, has a cone-like geometry and is located between 1 and 15 pc from the central source, which is much smaller than the observed extended emission cones in Seyfert 2 galaxies \citep{Kinkhabwala02,Guainazzi08}. One reason for this compactness could be that in Seyfert 2 galaxies, we can not observe the high density, high emissivity NLR gas that we observe here, since the torus blocks our view. So we are only seeing the lower density, low emissivity gas in Seyfert 2 galaxies, which is visible due to the very low continuum flux. From optical studies it is clear that the high-ionization as well as the high critical density lines tend to be stronger in Seyfert 1's, see e.g. \citet{Bennert06}. This is consistent with the compact size of the NLR we find in NGC 5548.

\section{Conclusions} 		\label{conclusions}

This is the first time that the long-term variability of the X-ray narrow line region has been studied in an AGN. NGC 5548 is the best studied AGN in the X-rays on a long time scale and has the best sampled \textit{RXTE} lightcurve which last 11 years. With this very rich set of data and the large change in flux after 2002, we are able to put a constraint on the location and geometry of the X-ray NLR. For NGC 5548 we favor a NLR which is located between 1 and 15 pc from the central source, is compact in the form of clouds or a narrow stream and has the geometry of an ionization cone. This is consistent with the picture sketched in Seyfert 2 galaxies, although the distance from the central source is much smaller by almost two orders of magnitude. The geometry and location of the NLR in Seyfert 1 galaxies can be further refined by a long-term monitoring program with regular (every few months) high resolution spectral observations of sufficiently variable sources. It will be interesting to study a larger sample of NLR emission lines in Seyfert 1 galaxies, to see if they all are of similar size or if NGC 5548 is unique with its compact NLR. 

\begin{acknowledgements}
SRON is supported financially by NWO, the Netherlands Organization for Scientific Research. R.G.D. would like to thank Elisa Costantini for discussion of this work.     
\end{acknowledgements}

\bibliographystyle{aa}
\bibliography{bibfiles}

\end{document}